\def\dl{\Bbb }  
\def\cala  {{\cal A}}

\def\caln  {{\cal N}}

\def\calr  {{\cal R}}

\def\zet           {{\dl Z}}

\newcommand\hsp[1] {\mbox{\hspace{#1 em}}}
\newcommand\nxt[1] {\\\raisebox{.12em}{\rule{.35em}{.35em}}\hsp{.6}#1}

\newcommand\sct[1]  {\mbox{$ $}\\[-.5em]{\bf #1.}\\[.16em]}
\def\alg           {algebra}

\def\be            {\begin{equation}}

\def\cft           {conformal field theory}
\def\cfts          {conformal field theories}
\def\chii          {\raisebox{.15em}{$\chi$}}
\def\cp            {Chan-Paton}

\def\ee            {\end{equation}}

\newcommand\erf[1] {(\ref{#1})}
\def\findim        {finite-dimensional}
\newcommand\Frac[2]{\mbox{\large$\frac{#1}{#2}$}}
\def\futnote#1     {\footnote{~#1}\ }

\def\hy            {$\mbox{-\hspace{-.66 mm}-}$}
\def\ii            {{\rm i}}

\def\irrep         {irreducible representation}

\long\def\labl#1   {\label{#1}\ee}

\def\onedim        {one-dimen\-sional}

\def\rep           {representation}
\def\resp          {respectively}

\def\su            {{\rm SU}(2)}

\def\twodim        {two-dimensional}

\def\wzwm          {WZW\hy model}
\def\wzwt          {WZW\hy theory}
\def\wzwts         {WZW\hy theories}

\documentclass[12pt]{article} \usepackage{amssymb,amsfonts}
\begin{document}

\begin{flushright}  {~} \\[-15 mm]  {\sf hep-th/9801191} \\[1mm]
{\sf CERN-TH/98-18} \\[1 mm]
{\sf January 1998} \end{flushright}
 
\begin{center} \vskip 15mm
{\Large\bf CLASSIFYING ALGEBRAS FOR BOUNDARY}\\[4mm]
{\Large\bf CONDITIONS AND TRACES ON SPACES OF}\\[4mm]
{\Large\bf CONFORMAL BLOCKS}\\[16mm]
{\large Christoph Schweigert} \\[3mm] CERN \\[.6mm] 
CH -- 1211~~Gen\`eve 23\\[11mm]
{\large J\"urgen Fuchs} \\[3mm]
Max-Planck-Institut f\"ur Mathematik\\[.6mm]
Gottfried-Claren-Str.\ 26, \  D -- 53225~~Bonn
\end{center}
\vskip 20mm
 
\begin{quote}{\bf Abstract}\\[1mm]
The boundary conditions of a non-trivial string background
are classified. To this end we need traces on various spaces of conformal
blocks, for which generalizations of the Verlinde formula are presented. 
\end{quote}
\vfill {}
\begin{flushleft}  {~} \\[-3 mm] {\sf CERN-TH/98-18} \\[1 mm]
{\sf January 1998} \\[3mm]
\noindent ------------------\\[1 mm]
{\footnotesize Slightly extended version of a talk given by C.\ Schweigert 
at the XXXI\,st International Symposium Ahrenshoop on the Theory of
Elementary Particles (Buckow, Germany, September 1997)}
\end{flushleft}

\sct{Introduction}
It has been known for a long time that the low energy effective action of 
superstring theories has solitonic solutions. Recently, it has become 
apparent that string perturbation theory in such a background can be
formulated in terms of world sheets with boundaries, where one imposes certain
non-trivial boundary conditions. Therefore, theories of open strings and 
conformal field theories on \twodim\ surfaces with boundaries have received
renewed interest. A central problem in these theories is the classification
of all consistent boundary conditions.  So far, however, most investigations 
have been limited either to models based on free world sheet theories or
on orbifolds of such theories, or to BPS-sectors of models
with extended supersymmetry. In this contribution we 
discuss  the structure of boundary conditions in an arbitrary rational \cft\ 
with a specific type of non-diagonal modular invariant. For a general
discussion of boundary conditions in \twodim\ \cft\ we refer to 
\cite{fuSc6}.
 
\sct{Modular invariants}
A chiral \cft\ typically admits several consistent torus partition functions.
Any non-trivial modular invariant of a rational \cft\ can be obtained 
\cite{mose2} by first extending the chiral algebra and then superposing an 
automorphism of the fusion rules. 
The extension of chiral algebras is by now fairly well understood, at least 
in the case of extensions by so-called simple currents \cite{fusS6,dolm,bant6},
and  can be described entirely in terms of a chiral half of the theory. As a 
consequence, such extensions do not raise any problem in the construction of 
open string theories that was not already encountered for closed strings so 
that we can assume that the modular invariant in question is of
pure automorphism type. Hence the modular invariant describes the pairing
between left-moving and right-moving fields, or, more precisely \cite{fuSc6}, 
the pairing between the two chiral \cfts\ on the oriented cover of the world 
sheet.

For simplicity, we also make a few more assumptions on the boundary conditions
we consider: first, we assume that the pairing for bulk fields is the same as 
in the case of closed orientable surfaces. In the terminology of Ref.\ 
\cite{fuSc6}, this means that we choose a trivial automorphism type, i.e.\ 
generalized Neumann boundary conditions. Next, we assume that the 
boundary preserves all symmetries of the bulk. Finally, we assume that we 
are dealing with one and the same chiral \cft\ on every type of \twodim\ 
surface. (This implies for instance \cite{fuSc6} that D-brane configurations 
with multiple wrapping are excluded.)

Under these assumptions, the possible boundary conditions have been classified
for a theory with the charge conjugation modular invariant \cite{card9}. 
A first investigation in the case of non-trivial modular invariants has been 
undertaken in \cite{prss3} for \wzwm s based on \su. 

The type of modular invariant we focus on generalizes the modular invariants 
that in the $A$-$D$-$E$ classification of \su\ modular invariants
are  of $D_{\rm odd}$-type. This modular invariant exists for level
$k=4l+2$ with $l$ integer:
\be Z(\tau)=\sum_{l=0}^{k/2} |\chii_{2l}^{}|^2+ \sum_{l=0}^{k/2} \chii_{2l+1}^{}
\chii_{k-2l-1}^* \, . \ee 
The full \cft\ described by this modular invariant can be regarded as a 
$\zet_2$-orbifold of the \wzwt\ on \su\ with the diagonal modular invariant. 
There are three types of primary fields:
\nxt Primary fields with integral isospin form the untwisted sector of the 
     orbifold. In the full theory, they are paired with themselves.
\nxt The twisted sector consists of the primary fields with half-integral 
     isospin. These fields $\Phi_l$ are paired 
     with some other primary field $\Phi_{k-l}$ which is obtained by taking
     the fusion product with the primary field $\Phi_k$ of highest possible 
     isospin, $\Phi_{k-l} = \Phi_k \star \Phi_l$. The primary field $\Phi_k$
     is a {\em simple current\/}: its fusion product with any other primary field 
     contains just one primary field with multiplicity one.
\nxt The twisted sector contains in particular the {\em fixed point\/}
     $\Phi_{k/2}$ which is mapped by the simple current  to itself, 
     $\Phi_k \star \Phi_{k/2} = \Phi_{k/2}$. In the twisted sector, we 
     distinguish between fixed points and non-fixed points.

In this note we consider \cfts\ with a similar $\zet_2$ symmetry: we assume 
that the theory contains a simple current $J$, i.e.\ a primary field such 
that its fusion product has the form $J\star \Phi_\Lambda = \Phi_{J\Lambda}$,
for any primary field $\Phi_\Lambda$. Moreover, we assume that $J$
squares to the vacuum primary field, $ J^2 = \Phi_0$, and that it has 
conformal weight $\Delta_J\in\zet+1/2$. Given such a simple current $J$, we
associate to every primary field $\Phi_\Lambda$ its {\em monodromy charge}
\be Q_J(\Lambda) := \Delta_\Lambda + \Delta_J - \Delta_{J\Lambda} \,\bmod \zet\ee
which generalizes the conjugacy class and is conserved in operator products.
One can show 
that in this situation the following expression gives a modular invariant
partition function.
\be 
Z= \sum_{Q(\Lambda)=0}\! \chii_\Lambda^{} \chii_{\Lambda^+}^* 
+ \!\sum_{Q(\Lambda)=1/2}\! \chii_\Lambda^{} \chii_{J \Lambda^+}^* \labl{44}
(This partition function looks like a $\zet_2$ orbifold of the original 
theory. It has a similar structure as the one of the type IIA
superstring in light cone gauge.) Again we have three types of primary 
fields: $N_0$ primary fields in the untwisted sector with monodromy charge
$Q(\Lambda)\,{=}\,0$. They all form orbits of length 2 under the action of $J$.
In the twisted sector, all fields have $Q(\Lambda)\,{=}\,1/2$. Among these
we have $N_1$ fields on full orbits and $N_f$ fixed points. 

\sct{Construction of the classifying algebra}
It has been shown in \cite{fuSc6} that a consistent boundary condition in
a \cft\ can be described by an automorphism of the fusion rules which 
preserves conformal weights, 
the {\em automorphism type\/} of the boundary condition, and a 
further degeneracy label, the {\em Chan-Paton type.} Moreover, it has been
shown \cite{fuSc6} that each Chan-Paton type corresponds to an \irrep\ of 
a so-called {\em classifying algebra}. This classifying algebra $\tilde\cala$   
has been computed in \cite{prss3} in the special case of \wzwts\
based on $\su$, using the explicit form of fusing matrices 
and operator product coefficients. It was observed that the possible \cp\ types
are in one-to-one correspondence to the orbits of the simple current $J$, 
where the fixed points of $J$ are counted with multiplicity two.
 
\sct{Traces on the space of conformal blocks}
The construction of the correct classifying algebra in the general case \erf{44}
requires the knowledge of some traces on the spaces of conformal blocks. 
In general, we consider the \findim\ vector space $B_{\vec\Lambda}$ of 
conformal blocks, where $\vec\Lambda=(\Lambda_1, \ldots,\Lambda_n)$ stands for 
a finite sequence of primary fields. In the case of three-point blocks
$B_{\lambda\mu\nu}$ its dimension is given by the Verlinde formula
\be \caln_{\lambda\mu\nu} = \dim B_{\lambda\mu\nu} =
\sum_\rho \, S_{\lambda\rho} S_{\mu\rho} S_{\nu\rho} \, / \,  S_{0\rho}\,. \ee

We now consider a collection $\vec J=(J_1, \ldots, J_n)$ of simple 
currents which fulfils $J_1\star J_2 \star\cdots\star J_n = 1$.
In this situation one can define a natural isomorphism between the spaces
of conformal blocks 
\be \Theta_{\vec J} \, : \quad 
B_{\vec\Lambda} \to B_{\vec J\vec \Lambda} \, , \ee
where we introduced the short hand $\vec J\vec \Lambda=(J_1\Lambda_1, \ldots,
J_n\Lambda_n)$. In the case of a simple current of order two, there are 
in particular isomorphisms:
\be \Theta_J\,: \quad B_{\lambda\mu\nu} \to B_{J\lambda \, J\mu \, \nu} \, . \ee
If $\lambda=f$ and $\mu=g$ are fixed points of $J$, then $\Theta_J$ is an 
{\em endomorphism\/} and we can consider the trace
\be \check \caln_{fg \nu} := {\rm Tr}_{B_{fg\nu}} \Theta_J
\, . \ee

The trace $\check\caln_{fg\nu}$ is an integer (surprisingly enough this is also
true for simple currents of any arbitrary order), and can be used to compute 
the dimension of the eigenspaces of $\Theta_J$ to the eigenvalues $\pm1$:
\be {\rm Tr}_{B_{fg\mu}} ( \Frac12 (1 \pm \Theta_J) )
= \Frac12 (\caln_{fg\mu} \pm \check\caln_{fg\mu}) \,, \labl8
which are manifestly non-negative integers.

These traces have already played an important role in chiral \cft, in the 
analysis of the twisted sector of extensions \cite{fusS6}. One finds that
there is a 
{\em fixed point theory\/} with modular matrix $\check S$ whose primary fields 
are in one-to-one correspondence with the fixed points. The traces are then 
given by a generalization of the Verlinde formula:
\be \check \caln_{fg \nu} = \sum_{h\; {\rm fix}} 
\,  \check S_{fh} \check S_{gh} S_{\nu h} \, / \, S_{0 h} \, . \ee

For \wzwts\  (and also for coset \cfts) the fixed point theories are obtained 
by folding Dynkin diagrams \cite{fusS3}. In general, it is conjectured that
they describe the modular properties of the one-point blocks on the torus with 
insertion the simple current $J$. 
The structure of fixed point theories is found in many 
places: it occurs in the twisted sector of extension modular invariants
\cite{fusS6}, in the very definition of coset \cfts\ \cite{fusS4}, 
in the Verlinde formula for WZW-theories based on non-simply connected Lie 
groups \cite{fusS6}
and in the topologically non-trivial components of the moduli spaces of 
holomorphic principal bundles with non-simply connected structure groups
over an elliptic curve \cite{schW3}. 

As a side remark we mention some other traces on the space of conformal
blocks which can be computed explicitly: on a four-point block on the
sphere with two identical primary fields one can consider the trace 
$Y_{i,i,j,k}$ of the permutation acting on the two identical insertions. 
Such traces appear in the theory of permutation orbifolds \cite{bohs}
as well as in the description of amplitudes on the M\"obius strip
\cite{prss}. Again, one has a generalization of the Verlinde formula:
\be Y_{i,i,j,k} = \sum_n P_{jn} P_{kn} S_{in} \, / \, S_{0n} \,, \ee
where we have introduced the matrix $P= T^{1/2} S T^2 S T^{1/2}$. 

\sct{The classifying algebra and its representation theory}
We are now in a position to display the classifying algebra $\tilde\cala$ 
\cite{fuSc5} for the \cft\ with torus partition function \erf{44}. The 
dimension of 
$\tilde\cala$ equals the number of bulk fields, $\dim\tilde\cala = N_0 + N_f$.
Moreover $\tilde\cala$ is $\zet_2$-graded and contains the fusion algebra of 
fields in the untwisted sector
as a subalgebra, since the operator products in the untwisted sector of an 
orbifold theory are the same as in the original theory. This leads to the 
following structure constants for $\tilde\cala$:
\be \tilde\caln_{\lambda\mu}^\nu = \left\{ \begin{array}{ll} 
\caln_{\lambda\mu}^\nu\quad & \mbox{if }\, Q(\lambda)=Q(\mu)=Q(\nu)=0 \\[.3em]
\check\caln_{\lambda\mu}^\nu\quad & \mbox{if there is precisely one 
full orbit} \\[.3em] 0&\mbox{else} \end{array}\right. \ee
One easily checks that this classifying algebra $\tilde\cala$ is commutative 
and associative, that $\Phi_0$ is a unit element and that the evaluation on 
the identity gives a conjugation on $\tilde\cala$:
$\check\caln_{fg}^0 = \delta_{f^+,g}$. As a consequence, $\tilde\cala$ is still
semi-simple, but in contrast to fusion algebras, its structure constants
can also be negative. Notice that $\tilde\cala$ is {\em not\/} a subalgebra of 
the fusion algebra $\cala$.

The representation theory of $\tilde\cala$ has the following structure: there
are $N_0+N_f = \frac12(N_0+N_1) + 2 N_f $ \irrep s, which are all \onedim. 
They are in correspondence to orbits of the simple current $J$: each of the 
orbits $\alpha$ of length two gives rise to a single \irrep\ $\calr_{(\alpha)}$:
\be \calr_{(\alpha)}(\Phi_\mu) = \left\{ \begin{array}{ll}
S_{\alpha\mu}\,/\,S_{0\alpha} \quad& \mbox{for}\quad Q(\mu)=0 \,, \\[.2em]
0&\mbox{for}\quad J\mu=\mu \,.
\end{array}\right. \ee
Each of the orbits $f$ of length one, the fixed points, gives rise to 
{\em two\/} different \irrep s $\calr_{(f+)}$ and $\calr_{(f-)}$:
\be \calr_{(f\pm)} (\Phi_\mu) = \left\{ \begin{array}{ll}
S_{f\mu}\, / \, S_{0f} \quad& \mbox{for}\quad Q(\mu)=0\,, \\[.2em]
\pm \check S_{f\mu}\, / \, S_{0f} &\mbox{for}\quad J\mu=\mu \,.
\end{array}\right. \ee
Notice that fixed points are fields in the twisted sector, and accordingly
the modular matrix $\check S$ of the fixed point theory appears.

\sct{The annulus amplitude, consistency checks}
To be able to perform several consistency checks, we compute the amplitude 
$ A_{ab}(t) = \sum_\mu A_{ab}^\mu \chii^{}_\mu(\frac{\ii t}2)$
for an annulus, where we impose boundary condition $a$ \resp\ $b$ on the two 
boundaries. We obtain the following result for the tensor $A^\mu_{ab}$:
\be \begin{array}{lll}
A_{\alpha\beta}^\mu = \caln_{\beta\mu}^\alpha + \caln_{\beta\mu}^{J\alpha}\,,&&
A_{\alpha (f\pm)}^\mu = \caln_{f\mu}^\alpha \,,  \\[.4em]
A_{(f\pm)(g\pm)}^\mu = \frac12( \caln_{f^+ g \mu}+\check\caln_{f^+ g \mu})\,,&&
A_{(f\pm)(g\mp)}^\mu = \frac12( \caln_{f^+ g \mu}-\check\caln_{f^+ g \mu}) \,.
\end{array}\ee

The tensor $A^\mu_{ab}$ allows to perform the following checks. We first
remark that all $A_{ab}^\mu$ are non-negative integers as it befits for an 
expansion of an open string partition function. This result is particularly 
non-trivial for fixed points, where it follows from \erf8. Next we observe
that the fact that the conjugation on the classifying algebra that is
the evaluation on the identity implies that the multiplicities of the vacuum
in the open string partition functions are either $0$ or $1$. This was a
consistency requirement in \cite{card9}. Finally we can check the consistency
of several factorizations. In the case of a sphere with four boundary circles 
with boundary conditions $a,b,c$ and $d$ we have 
$\sum_\mu A_{ab}^\mu A_{cd}^{\mu^+} = \sum_\mu A_{ac^+}^\mu A_{b^+d}^{\mu^+}$;
also, $A^\mu A^\nu = \sum_\lambda \caln_{\mu\nu}^\lambda A^\lambda$, which
gives the correct factorization of the annulus.
We also remark that the heuristic argument used in \cite{card9} to derive the
classifying algebra for the charge conjugation modular invariant can be
generalized to the case of our interest; for details we refer to \cite{fuSc5}.

\sct{Conclusions}
The structure of the classifying algebra $\tilde\cala$ is actually closely 
related to the fusion \alg\ of another type of modular invariants, namely 
those of `$D_{\rm even}$-type' (which are also known as integer spin simple 
current extensions). In particular, our results for the classifying algebra look
as if the boundary theory were extended by the {\em half\/}-integer spin 
simple current $J$.

This is indeed most remarkable, because in the case of extensions, this
structure is a consequence of the powerful consistency requirements of 
modular invariance.  But for the crosscap as well as for the annulus and
the M\"obius strip, there is no analogue of a modular group. In string
theory it is usually argued that tadpole cancellation provides a substitute
for such consistency conditions. Note, however, that for our investigations
we did not have to assume that the \cft\ is part of
a string compactification (e.g., the central charge is not restricted), so that
the conditions of tadpole cancellation cannot even be formulated.
Still, it seems that already on a pure \cft\ level there are similar powerful
constraints; to unravel the underlying structure seems to be a promising task.

Finally we mention that the construction of a classifying algebra for
modular invariants of automorphism type that are not simple current 
automorphisms is still an open problem. A particularly interesting case 
is the one of generalized
Neumann boundary conditions for the true diagonal modular
invariant, where the relevant automorphism is just charge conjugation.

 \def\wb{\,\linebreak[0]} \def\wB {$\,$\wb}
 \def\Bi{\bibitem }
 \newcommand\Erra[3]  {\,[{\em ibid.}\ {#1} ({#2}) {#3}, {\em Erratum}]}
 \newcommand\BOOK[4]  {{\em #1\/} ({#2}, {#3} {#4})}
 \newcommand\J[5]   {{\sl #5}, {#1} {#2} ({#3}) {#4} }
 \newcommand\Prep[2]  {{\sl #2}, preprint {#1}}
 \def\comp  {Com\-mun.\wb Math.\wb Phys.}
 \def\nupb  {Nucl.\wb Phys.\ B}
 \def\phlb  {Phys.\wb Lett.\ B}

\end{document}